\documentclass[sigconf,nonacm]{acmart}
\AtBeginDocument{%
  }

\usepackage{subcaption}
\usepackage{xspace}

\newcommand{\capybara}{\textsc{Capybara}\xspace}
\begin{document}

\title{How Children Collaborate within Programmable AR Environments with Co-Located Collaborative Features}

\author{Romina Mahinpei}
\email{rmahinpei@princeton.edu}
\orcid{0000-0002-7500-5928}
\affiliation{%
  \institution{Princeton University}
  \country{United States}
}

\author{Diya Ajay Hundiwala}
\email{dh3163@princeton.edu}
\orcid{0009-0001-6927-1103}
\affiliation{%
  \institution{Princeton University}
  \country{United States}
}

\author{Sandy Zhang}
\email{sz4476@princeton.edu}
\orcid{0009-0001-1289-7033}
\affiliation{%
  \institution{Princeton University}
  \country{United States}
}

\author{Lana Glisic}
\email{lglisic@princeton.edu}
\orcid{0009-0005-3342-0663}
\affiliation{%
  \institution{Princeton University}
  \country{United States}
}

\author{Andrés Monroy-Hernández}
\email{andresmh@princeton.edu}
\orcid{0000-0003-4889-9484}
\affiliation{%
  \institution{Princeton University}
  \country{United States}
}

\renewcommand{\shortauthors}{Mahinpei et al.}

\begin{abstract}
Programmable augmented reality (AR) environments are emerging as a promising way to support children’s creative learning through embodied interaction with digital characters and physical space. At the same time, AR systems are increasingly capable of supporting co-located collaborative experiences. However, little is known about how children collaborate within programmable AR environments offering co-located collaborative features. In response, we extended Capybara, an existing programmable AR application for children, with co-located collaborative features supporting shared visibility and interaction across devices. We then conducted workshops with 9 children to examine whether and how collaboration emerges during use. Across our workshops, collaboration was often lightweight and implicit, emerging through three complementary forms: parallel play with social awareness, iterative remixing, and spontaneous peer support. Together, our findings provide insights for designing future child-centered programmable AR systems that better support co-located collaborative experiences.
\end{abstract}




\begin{teaserfigure}
    \centering
    \begin{subfigure}{0.33\textwidth}
        \centering
        \includegraphics[width=0.94\textwidth]{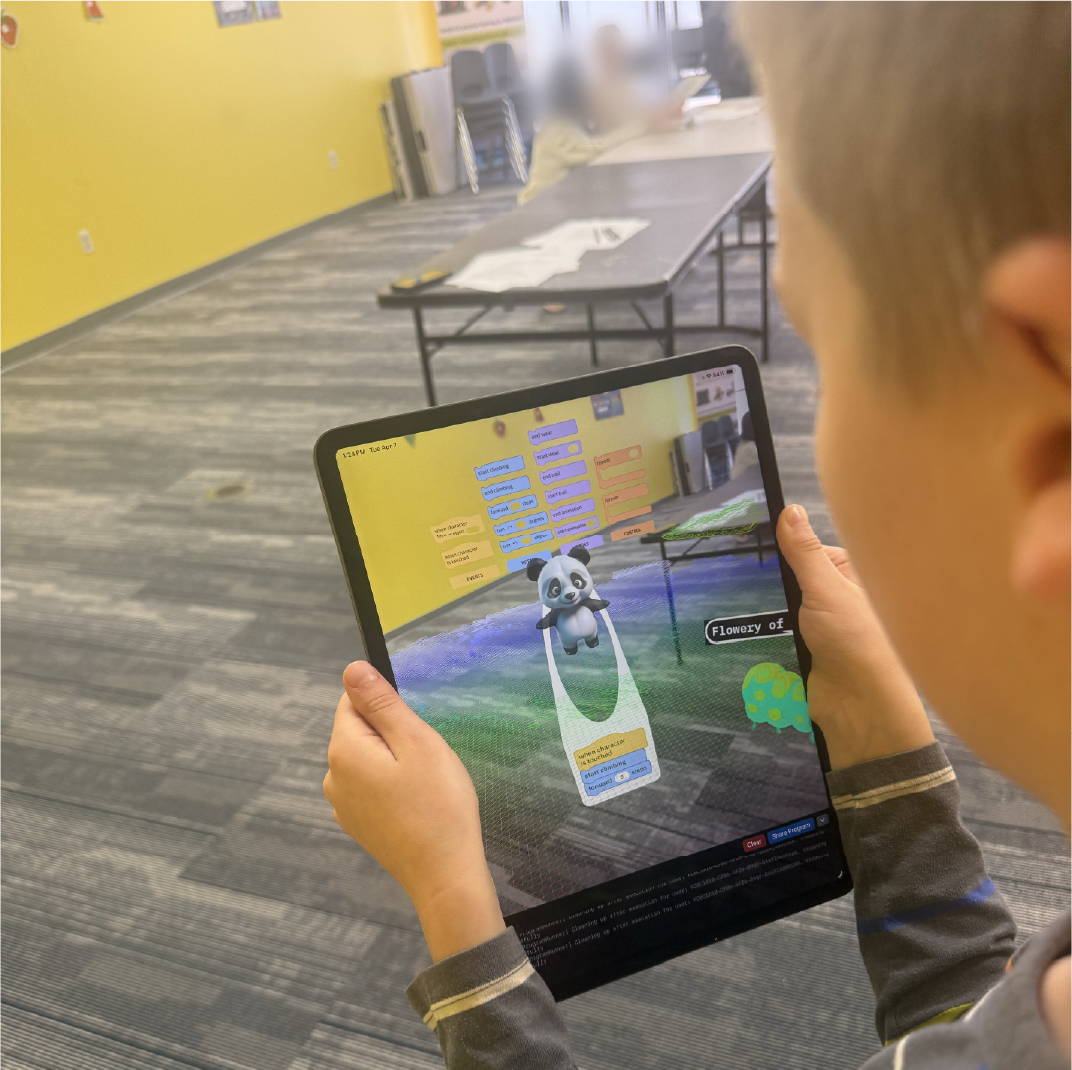}
        \caption{\textbf{Parallel play with social awareness}: A child plays with their panda character while seeing their peer's caterpillar character.}
        \label{fig:teaser_play}
    \end{subfigure}
    \hfill
    \begin{subfigure}{0.31\textwidth}
        \centering
        \includegraphics[width=\textwidth]{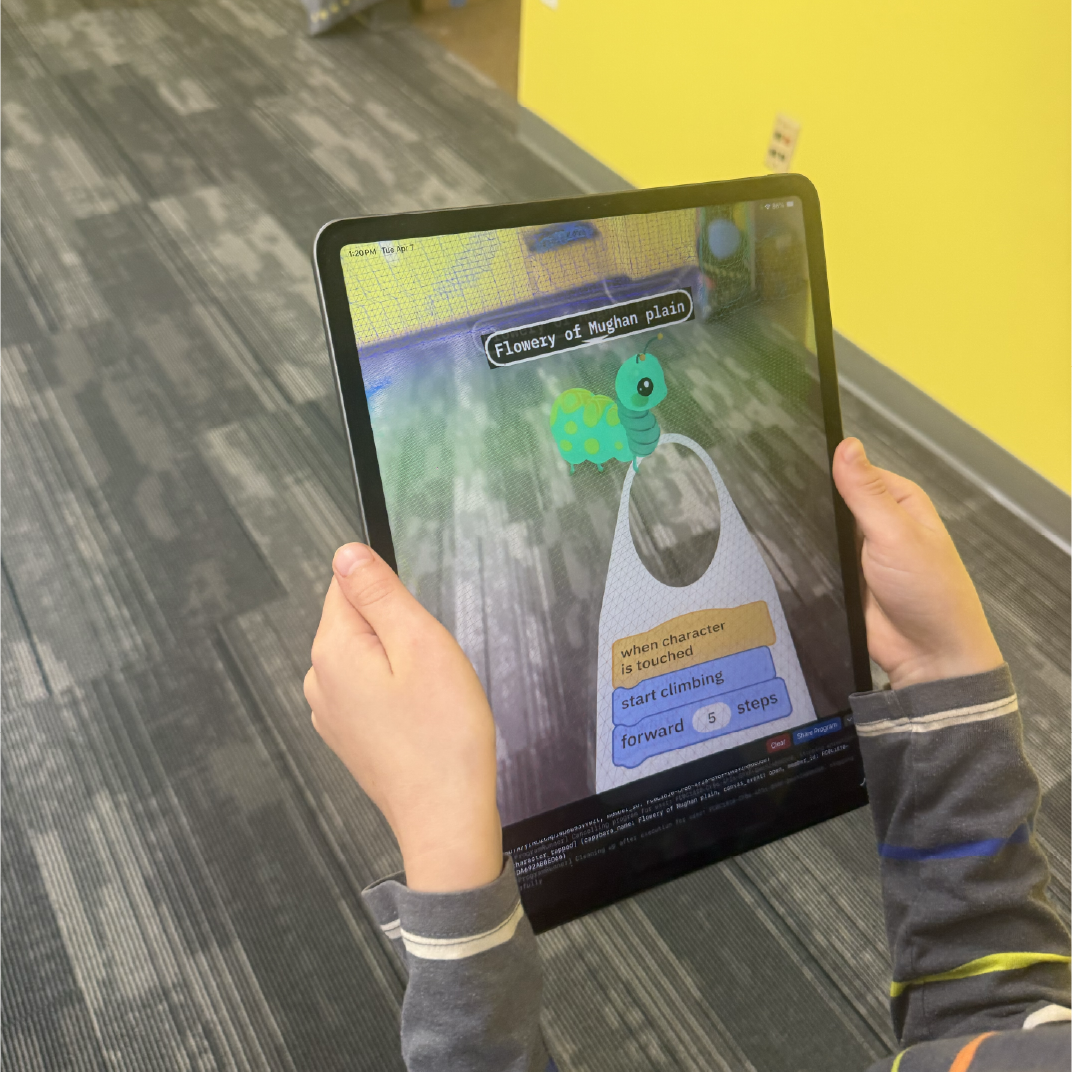}
        \caption{\textbf{Iterative remixing}: A child views the block-based program of their peer and copies it in order to modify the program.}
        \label{fig:teaser_share}
    \end{subfigure}
    \hfill
    \begin{subfigure}{0.31\textwidth}
        \centering
        \includegraphics[width=\textwidth]{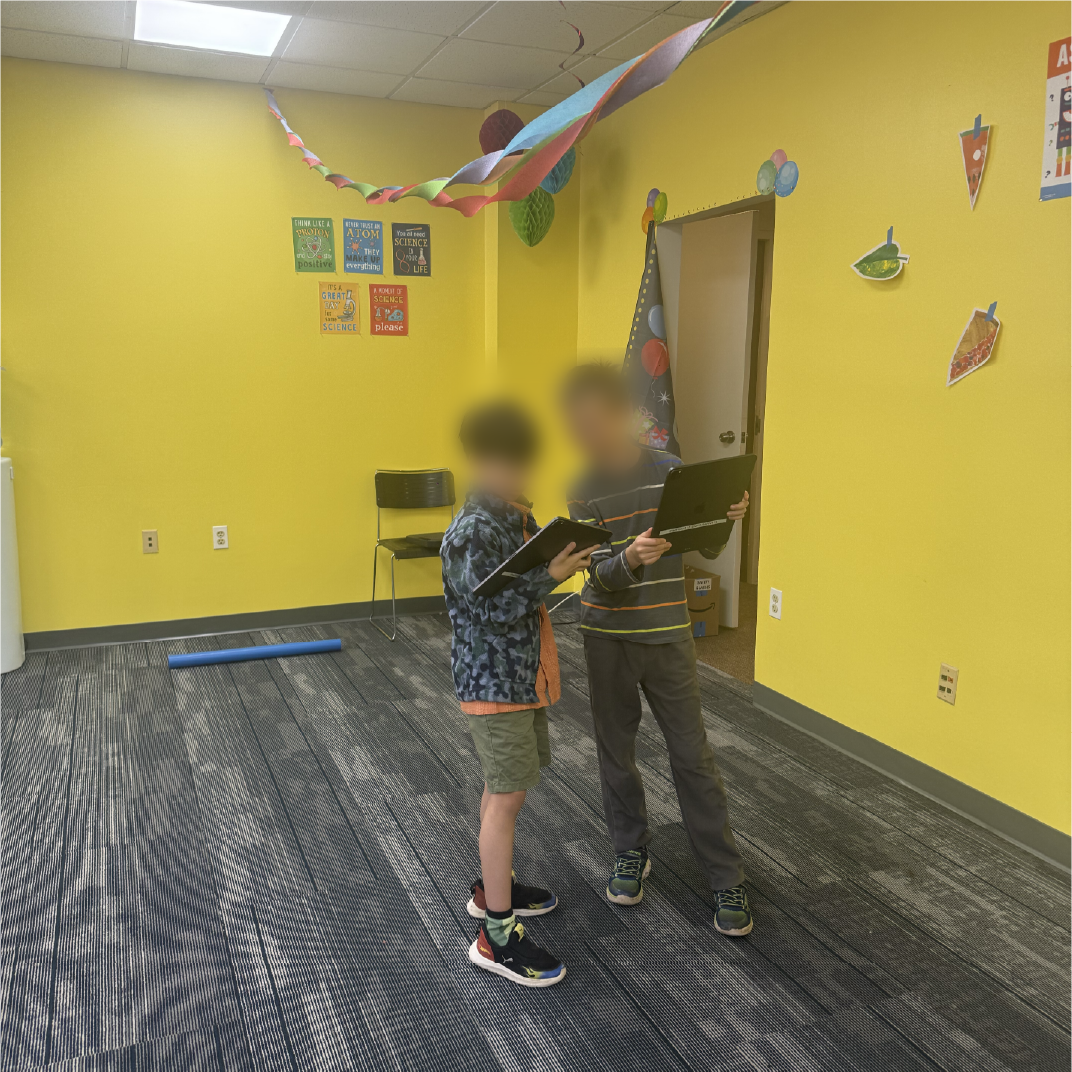}
        \caption{\textbf{Spontaneous peer support}: Two children help each other with using the various features offered within the AR environment.}
        \label{fig:teaser_help}
    \end{subfigure}
    \caption{\textbf{Collaboration within programmable AR with co-located collaborative features}: Across our workshops, collaboration was often implicit and present in three complementary forms: (a) \textbf{parallel play with social awareness}: children pursued individual goals while remaining aware of the activity of peers; (b) \textbf{iterative remixing}: children shared, borrowed, and modified each other’s code and ideas; and (c) \textbf{spontaneous peer support}: children provided spontaneous support when needed by peers.}
    \label{fig:teaser_figure}
\end{teaserfigure}

\maketitle

\begin{center}
\fbox{%
  \parbox{0.9\linewidth}{%
    \centering
    \textbf{Work published at CSCW 2026. Please cite the published version accordingly.}
  }%
}
\end{center}

\newpage
\section{Introduction}
Children often learn, create, and play alongside peers, with prior work having shown that idea exchange, mutual inspiration, and feedback can lead to richer creative outcomes and deeper engagement~\cite{Bruckman98, Hernandez07, Brennan10, Resnick17}. In parallel, programmable augmented reality (AR) environments are emerging as a promising way to introduce children to computing and support creative expression through embodied interaction with digital characters and physical space~\cite{Liaqat25, Lei25, Alves25, Chung25}, with many AR systems also becoming increasingly capable of supporting co-located collaborative experiences in which multiple users share and interact within the same augmented environment~\cite{Guo21, Poretski21, Dagan22, Petrov23, Numan25, Wang25}. 

Together, these developments point toward a new category of child-centered computing experiences: \textit{programmable AR environments with co-located collaborative features}. However, little is known about whether and how collaboration emerges in such settings, where interactions involve not only coordination across physical space but also across digital characters, code blocks, and programs. Understanding these behaviors is important for designing future child-centered programmable AR systems that better support collaborative learning, creativity, and play. Accordingly, we ask: \textit{When children use programmable AR environments with co-located collaborative features together in the same physical space, what forms of collaboration emerge (if any)?} 

In response, we extended \capybara~\cite{Lei25}, an existing iOS application that offers a programmable AR environment for children, with co-located collaborative features that support shared visibility and interaction across devices. We then conducted 90-minute workshops with 9 children, split into 4 groups, to examine whether and how collaboration emerges during use. Across our workshops, collaboration was often lightweight and present in three complementary forms: parallel play with social awareness, iterative remixing, and spontaneous peer support. Our findings suggest that collaboration in programmable AR environments may emerge implicitly rather than through explicit coordination, providing insights for the design of future child-centered programmable AR environments that better support co-located collaborative experiences.

\section{Related Work}
Children often learn, create, and play through collaboration with peers, with prior work showing that peer interaction through idea exchange, remixing, and feedback supports deeper engagement and richer creative outcomes~\cite{Bruckman98, Brennan10, Resnick17}. 
Reflecting these insights, many creative platforms for children have incorporated collaborative features to scaffold peer interaction. A prominent example is Scratch’s online community, where young creators share, remix, and build upon one another’s projects, enabling new forms of participation and expanding creative possibilities~\cite{Hernandez07}.

Building on this foundation, prior work has explored AR as a medium for children’s creativity and learning. Systems such as StoryMakAR~\cite{Terrell20} and \capybara~\cite{Lei25} use block-based programming to help children author interactive AR experiences, while other projects show how AR can support storytelling~\cite{Zhang24, Alves25} and creative writing~\cite{Chung25}. Recent work further finds that young people are particularly interested in creating playful, character-driven, and socially meaningful AR experiences~\cite{Liaqat25}, highlighting AR’s potential to position children as active creators rather than passive consumers of technology. At the same time, a complementary line of work examines how AR can support co-located collaboration through shared spaces, embodied coordination, and playful interaction. Systems such as Blocks~\cite{Guo21} and Dream Garden~\cite{Petrov23} enable users to co-create persistent AR content, while Project IRL~\cite{Dagan22} and Wandering Spirit~\cite{Wang25} show how AR can foster in-person play and coordination through shared virtual or physical objects. Other projects similarly highlight how physical space can anchor collaboration and support joint activity~\cite{Poretski21, Numan25}.

While prior work has studied AR for children and co-located collaboration within AR environments separately, little is known about whether and how children collaborate in programmable AR environments that offer co-located collaboration, in which interaction spans code, digital characters, and physical space. We address this gap by examining the forms of collaboration that emerge when children use such environments together in a shared physical space.

\section{System Overview}
\begin{figure}[htbp]
    \centering
    \begin{subfigure}[t]{0.23\textwidth}
        \centering
        \includegraphics[height=5cm]{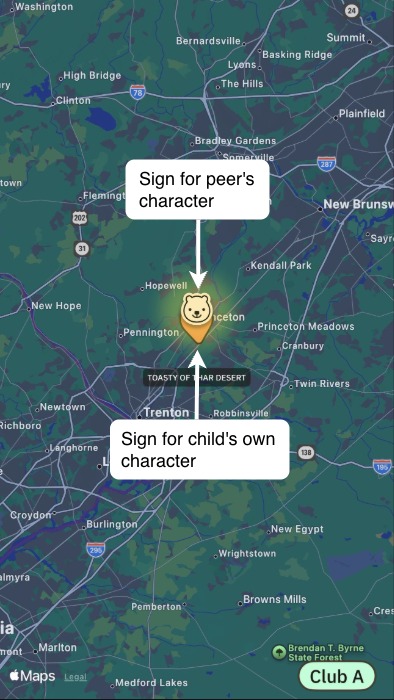}
        \caption{\textbf{Shared map}: Children can see who is present in the club and where they are located.}
        \label{fig:shared_map}
    \end{subfigure}
    \hfill
    \begin{subfigure}[t]{0.23\textwidth}
        \centering
        \includegraphics[height=5cm]{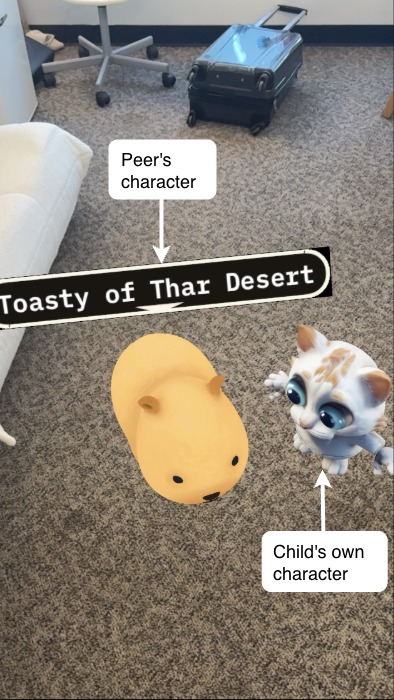}
        \caption{\textbf{Peer interaction}: Children can see, move, and activate the characters of peers.}
        \label{fig:peer}
    \end{subfigure}
    \hfill
    \begin{subfigure}[t]{0.23\textwidth}
        \centering
        \includegraphics[height=5cm]{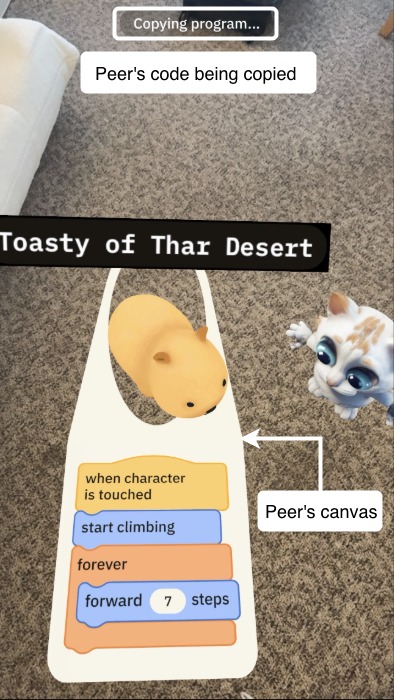}
        \caption{\textbf{Code remixing (i)}: Children can copy code directly from a peer's canvas.}
        \label{fig:remixing_copy}
    \end{subfigure}
    \hfill
    \begin{subfigure}[t]{0.23\textwidth}
        \centering
        \includegraphics[height=5cm]{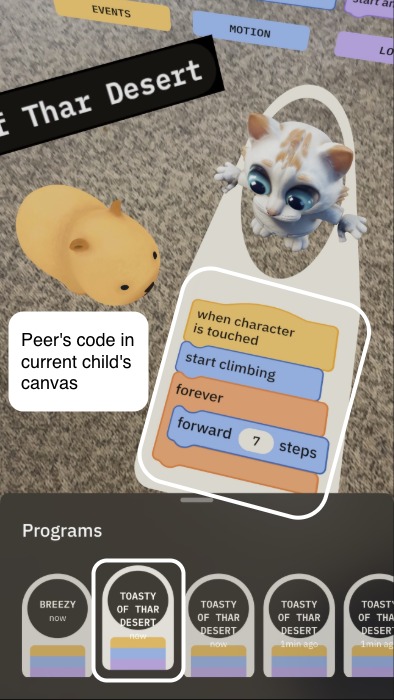}
        \caption{\textbf{Code remixing (ii)}: The copied code is then available on the child's canvas for remixing.}
        \label{fig:remixing_own}
    \end{subfigure}
    \caption{\textbf{Co-Located collaborative features added to \capybara}: We extended the original version of \capybara with three distinct features to support shared awareness, peer interaction, and code remixing.}
\label{fig:collab_features}
\end{figure}

\capybara is an AR-based iOS application designed for tablets and phones that enables children to create, customize, and program interactive 3D characters in the physical world through a block-based programming interface and features powered by generative artificial intelligence (genAI)~\cite{Lei25}. In the original system, learners could customize their 3D characters using genAI, animate them through body-based interactions, and define behaviors using drag-and-drop code blocks that connect virtual elements to the physical environment (see Appendix~\ref{app:capybara_features} for a gallery of these features).

For this work, we extended \capybara with co-located collaborative features that introduce shared awareness, peer interaction, and code remixing into the AR-based programming environment. Specifically, our features were informed by prior research showing that young people value social connection, shared presence, and opportunities to engage with others’ creations in AR environments~\cite{Liaqat25}. To use our extended version of \capybara, children join a shared ``club" on the app, which then allows them to see and interact with other members within that club.

Within each club, we introduced a set of real-time awareness and interaction mechanisms that make other users visible and interactive. First, children can view all members of their club on a shared map, providing a global overview of who is present and where activity is happening (\textbf{Figure~\ref{fig:shared_map}}). Second, children can see each other’s characters co-existing in the same AR environment, creating a sense of co-presence, and can also directly interact with each other's characters by moving them (\textbf{Figure~\ref{fig:peer}}). Third, children can share code by directly copying from each other's canvas and even activate each other's characters by physically touching the virtual character (\textbf{Figures~\ref{fig:remixing_copy},\ref{fig:remixing_own}}). 

These interactions are designed around the notion of \textit{multiple perspectives}. While children can see and manipulate shared elements (e.g., each other's characters), their views are not fully synchronized. For instance, if child A moves child B’s character, this change is not reflected in child B’s view. This intentional asymmetry preserves individual ownership while still enabling collaborative exploration, aligning with prior findings that differences in perspective can create opportunities for engagement and learning in AR environments~\cite{Liaqat25}.

To summarize our co-located collaborative features, we describe our system through a vignette. Two children, Aisha and Mateo, join the same club. On the map, Aisha notices Mateo nearby and walks over to him. Through her device, she sees Mateo’s character placed next to her own in the AR environment. Curious, Aisha drags Mateo’s character across the room to explore how it behaves. On Mateo’s screen, however, his character remains in its original position, allowing him to continue building without interruption. Meanwhile, Mateo activates Aisha’s character to observe its behavior and copies part of her code to use in his own program. Through these interactions, both children engage with each other’s ideas while maintaining individual creative control.


\section{Workshop Design}
\subsection{Participants} 
After receiving approval from our Institution's Review Board, we recruited participants in collaboration with a local after-school center that offers programs for children ages 5-12. Families who had participated within the organization's past programs were contacted and informed of our workshop, which were offered at two time slots. Across the two time slots, we recruited a total of 9 children (ages 5-11; 5 girls, 4 boys) who were assigned to groups of 2-3  (see  Appendix~\ref{app:participant_table} for participant demographics). Each family received a \$30 gift card as a token of appreciation for their time.

\subsection{Workshop Protocol} 
After parents and children completed our consent and assent forms, the workshop began with a brief introduction and a walkthrough of \capybara's original features and our collaborative features. The session was then structured around two stages, during which each group joined their specific club instance such that only members in that group could view and interact with each other's characters. Importantly, each group also consisted of at least two researchers, with one researcher executing the protocol and the other taking observational notes. The workshop itself consisted of (1) a structured storytelling activity based on \textit{The Very Hungry Caterpillar}, during which children recreated and extended the story using Capybara, and (2) an open-ended exploration period. Workshops concluded with a brief focus group on the children's overall experience as well as a background survey (see Appendix~\ref{app:workshop_details} for our focus group questions and background survey).


\subsection{Data Collection \& Analysis} 
We collected three forms of data during the workshops: (1) observational notes recorded by a researcher for each group using a shared note-taking template that prompted observers to document what children did, whether and how they collaborated, and notable interactions or behaviors, (2) audio recordings of each group that were automatically transcribed and checked by one author for accuracy, and (3) post-workshop reflections from all researchers. We analyzed the data using an inductive open coding approach. Two authors independently conducted line-by-line coding across the full dataset to identify emergent forms of collaboration. They then met to reconcile their codes and used affinity diagramming to organize recurring patterns into higher-level themes concerning collaboration, which were further discussed and refined with the broader research team who conducted and observed the workshops.

\section{Results}
Throughout our workshops, collaboration was often implicit rather than explicitly coordinated, emerging through three lightweight, complementary forms: parallel play with social awareness, iterative remixing, and spontaneous peer support.


\subsection{Parallel Play with Social Awareness}
Rather than directly collaborating on a shared task, children often observed, reacted to, and drew inspiration from one another’s creations and interactions while pursuing individual goals and narratives. Specifically, children’s creations and behaviors evolved in response to what peers were doing nearby. For example, one observer noted that after P3 started a virtual trail with their character, P4 also looked into the trail feature. P5 also noted how P6’s character changed in response to what they were doing: 
\begin{quote}
    \textbf{P5}: \textit{``Yeah, when my capybara turned into a tiger, yours also turned into a tiger."}
\end{quote}

\noindent Similarly, after seeing the characters of their peers, P9 modified their play narrative in response:
\begin{quote}
    \textbf{P9}: \textit{``It was pretty cool how we went through your two characters; we were huge!"}
\end{quote}

These interactions often resembled forms of parallel play but with strong social awareness and inspiration between participants. The extent to which children drew inspiration from others appeared to vary based on age and prior AR/VR experience. Younger children, in particular, looked to older peers for cues about what was possible within the environment, as was the case with younger participants P2 and P6 who relied on P1 and P5, respectively, for ideas.

\subsection{Iterative Remixing}
Children also engaged in iterative remixing by copying and modifying one another’s block-based programs. Rather than treating programs as fixed artifacts owned by a single creator, children frequently experimented with each other’s code in playful ways. Remixing was often framed playfully by children themselves, who described the code sharing feature as``stealing each other’s code'' or ``sharing a secret.'' In such cases, remixing became a playful mechanism for experimentation through which children iteratively transformed one another’s programs into new narratives: 
\begin{quote}
    \textbf{P8}: \textit{``I stole your code and turned them and made them different!''}
\end{quote}

\noindent Interestingly, collaboration did not always occur synchronously, with some children expressed enjoyment in creating programs that could later ``surprise'' their peers.

\subsection{Spontaneous Peer Support}
A third form of collaboration involved spontaneous peer support during moments of uncertainty or discovery. Although researchers were present throughout the workshops to provide support, children frequently turned to one another for assistance. Specifically, help often emerged organically in response to immediate needs rather than through formally assigned roles. In several cases, children who had discovered a feature showed it to peers who became interested in the feature after observing their peer. For example, P4 helped P3 start a virtual trail for their character while P3, who was more familiar with the concept of ``degrees,'' helped P4 get their character to turn by a specified number of degrees. Similarly, P9 who had re-discovered that they could activate the characters of others, informed others about this capability.    

\section{Discussion}
Our findings suggest that collaboration within programmable AR environments that offer co-located collaborative features may not always take the form of explicit collaboration toward a shared objective. Even during our structured storytelling activity, children engaged in more lightweight and implicit forms of collaboration, including observing peers, remixing ideas, and providing spontaneous peer support. As such, rather than enforcing explicit collaboration, programmable AR systems may benefit from supporting collaboration that emerges organically through shared awareness, curiosity, and playful interaction.

One possible explanation for the prevalence of implicit collaboration is that children appeared to value preserving individual perspectives alongside social interaction. Participants often pursued personal goals and narratives while still drawing inspiration from and engaging with the creations of peers. In this sense, collaboration became intertwined with self-expression rather than replacing it. The intentionally asymmetric nature of our system, in which children could interact with shared elements without fully synchronizing perspectives, may have helped preserve this sense of individual authorship while still enabling lightweight social interaction.

More broadly, our findings complement established theories of children's play and learning. The lightweight forms of collaboration we observed resemble classic accounts of parallel and associative play, in which children engage in their own activities while remaining socially aware of others~\cite{Parten32} while iterative remixing extends prior work on creative learning and remixing in programming contexts~\cite{Resnick17}. Together, these connections suggest that programmable AR may provide a new medium through which well-established collaborative learning processes emerge, rather than introducing entirely new forms of collaboration.

\section{Limitations \& Future Work}
Our findings are based on workshops with a relatively small group of participants, and future work should examine whether our findings generalize across larger and more diverse populations of children. While our workshops included children spanning a relatively broad age range, they were not designed to examine developmental differences systematically. Future work could investigate how age, prior programming experience, and other learner characteristics influence collaborative behaviors and interactions with programmable AR. Additionally, our system incorporated a specific set of collaborative features centered around shared awareness, peer interaction, and code remixing. Although our observations suggest these features supported collaboration, our study was not designed to isolate their individual contributions. Future work could compare alternative collaborative designs, such as different mechanisms for synchronization, to better understand how specific features shape collaboration within programmable AR environments.

\begin{acks}
This work was made possible through the funding provided by the Princeton University's NextG Innovation Grant.    
\end{acks}

\newpage
\bibliographystyle{ACM-Reference-Format}
\bibliography{references}

\appendix
\section{Capybara Features}
\label{app:capybara_features}
\begin{figure*}[t]
    \centering
    \includegraphics[width=\textwidth]{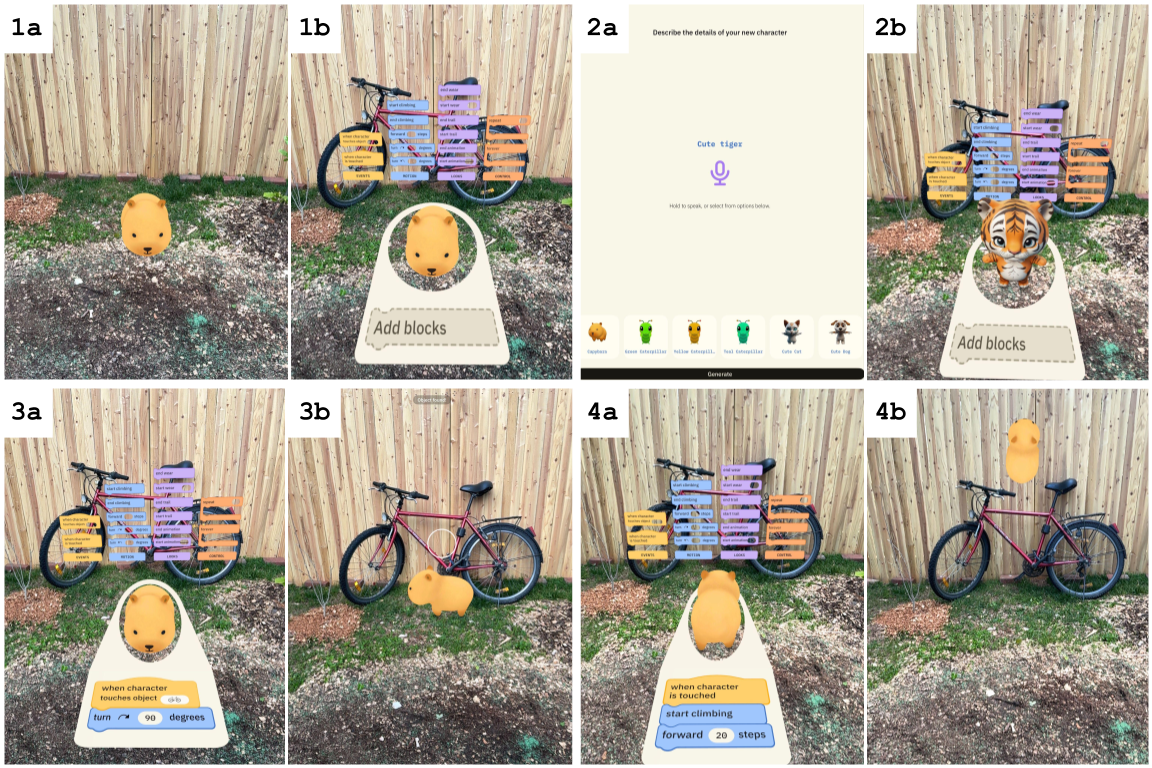}
    \caption{\textbf{Features of the Capybara application}. (1a) Once the application is started, the Capybara character is anchored to the ground in front of the user; (1b) pressing on the character opens a canvas consisting of code blocks. (2a) The user specifies a new character model via a spoken prompt; (2b) the character transforms into an AI-generated creation based on the prompt (e.g., a tiger). (3a) The tiger is programmed to turn right when it touches a real-world bicycle; (3b) the tiger turns right upon contact with the bicycle. (4a) The tiger is programmed to climb for 20 steps; (4b) the tiger climbs up the real-world fence.}
    \label{fig:feature_gallery}
\end{figure*}

We provide a gallery of the main features offered within the original version of Capybara in Figure~\ref{fig:feature_gallery}. This includes block-based programming, GenAI character customization, object and wall detection alongside climbing.

\label{app:features_appendix}

\section{Participant Demographics}
\label{app:participant_table}
We provide the demographics of the children who participated in our workshops in Table~\ref{tab:participants}.

\begin{table*}[t]
\centering
\footnotesize
\caption{\textbf{Workshop Participants}: Demographics of children who participated in our workshops.}
\label{tab:participants}
\begin{tabular}{cclccc}
\toprule
\textbf{Group \#} & \textbf{Participant \#} & \textbf{Age} & \textbf{Gender} & \textbf{Has Coding Experience?} & \textbf{Has AR/VR Experience?} \\
\midrule
1 & 1 & 11 & Female & Yes & No  \\ 
1 & 2 & 7  & Female & Yes & No  \\
\hline
2 & 3 & 7  & Male   & Yes & No  \\
2 & 4 & 9  & Male   & Yes & Yes \\
\hline
3 & 5 & 9  & Female & No  & Yes \\
3 & 6 & 5  & Male   & No  & No  \\
\hline
4 & 7 & 8  & Female & No  & No  \\
4 & 8 & 10 & Female & Yes & Yes \\
4 & 9 & 8  & Male   & No  & Yes \\
\bottomrule
\end{tabular}
\end{table*}

\section{Workshop Questions \& Survey}
\label{app:workshop_details}
We include the interview questions and background survey used in our workshops below.

\subsection{Semi-Structured Interview Questions}
\begin{itemize}
    \item How did you find recreating the story of the ``The Very Hungry Caterpillar''? 
    \begin{itemize}
        \item What did you like and not like about it?
    \end{itemize}
    \item How did you find copying and sharing code with your friend?
    \begin{itemize}
        \item What did you like and not like about it?
    \end{itemize}
    \item Did it feel like you were creating with others or on your own? \item Did others’ work or code influence what you did?
    \item What was your favorite feature in the app?
    \item What was your favorite part of the workshop?
    \item How would you describe your overall experience today?
\end{itemize}

\subsection{Background Survey}
\begin{enumerate}
    \item What is your age (in years)? 
    \item What is your gender identity?
    \item Do you have any coding experience? If yes, please provide 1-2 examples of platforms you have used for coding (e.g., Scratch).
    \begin{itemize}
        \item [] Yes
        \item [] No
    \end{itemize}
    \item Have you worked with augmented/virtual reality (AR/VR) before? If yes, please provide 1-2 examples of AR/VR you have used (e.g., Pokémon Go).
    \begin{itemize}
        \item [] Yes
        \item [] No
    \end{itemize}
\end{enumerate}

\end{document}